\documentclass[aps,prd,preprintnumbers,nofootinbib,twocolumn]{revtex4-2}
\usepackage{bm}
\usepackage{latexsym}
\usepackage{dcolumn}
\usepackage{amsmath,amsfonts,amssymb}
\usepackage{graphicx,float}
\usepackage{appendix}
\usepackage[hyperfootnotes=true]{hyperref}
\usepackage{mciteplus}
\usepackage{subcaption}
\usepackage{caption}
\usepackage{bbold}
\usepackage{slashed}
\usepackage{mathbbol}
\DeclareSymbolFontAlphabet{\amsmathbb}{AMSb}%
\usepackage{amsthm}
\usepackage{hyperref}
\usepackage{color}
\usepackage{mathrsfs}
\usepackage{setspace}
\usepackage{braket}
\usepackage{orcidlink}

\DeclareMathAlphabet{\mathpzc}{OT1}{pzc}{m}{it}
\usepackage{calligra}
\DeclareMathAlphabet{\mathcalligra}{T1}{calligra}{m}{n}
\DeclareFontShape{T1}{calligra}{m}{n}{<->s*[2.2]callig15}{}

\def\be {\begin{equation}}
\def\ee {\end{equation}}
\def\bea {\begin{eqnarray}}
\def\eea {\end{eqnarray}}
\def\bc {\begin{center}}
\def\ec {\end{center}}
\def\bfg {\begin{figure}}
\def\efg {\end{figure}}
\def\bi {\begin{itemize}}
\def\ei {\end{itemize}}

\def\beq{\begin{equation}}
\def\eeq{\end{equation}}
\def\br{\begin{eqnarray}}
\def\er{\end{eqnarray}}
\newcommand{\eel}[1] {\label{#1}\end{equation}}

\emergencystretch=\maxdimen
\newcommand{\bdm}{\begin{displaymath}}
\newcommand{\edm}{\end{displaymath}}

\begin{document}

\title{Theoretical and Observational Implications of  Planck's Constant as a Running Fine Structure Constant}

\author{Ahmed Farag {\bf{A}}li $^\nabla$$^{\triangle}$\orcidlink{0000-0001-8944-6356}}
\email[email: ]{aali29@essex.edu ; ahmed.ali@fsc.bu.edu.eg}

\author{Jonas {\bf{M}}ureika$^{\square, \bigcirc \orcidlink{0000-0002-9926-4465}}$}
\email[email: ]{jmureika@lmu.edu}
\author{Elias~C.~{\bf{V}}agenas$^{\oplus}$\orcidlink{0000-0002-8739-9506}}
\email[email: ]{elias.vagenas@ku.edu.kw}

\author{Ibrahim {\bf{E}}lmashad $^{\triangle}$ \orcidlink{0000-0002-1554-5359}}
\email[email: ]{ibrahim.elmashad@fsc.bu.edu.eg}

\affiliation{ $^{\triangle}$ Department of Physics, Benha University, Benha 13518, Egypt}
\affiliation{$^{\nabla}$Essex County College, 303 University Ave, Newark, NJ, USA 07102}
\affiliation{ $^{\square}$ Department of Physics, Loyola Marymount University, 1 LMU Drive, Los Angeles, CA, USA 90045}
\affiliation{ $^{\bigcirc}$ Kavli Institute for Theoretical Physics, University of California, Santa Barbara, CA, USA 93106}
\affiliation{$^{\oplus}$ Department of Physics, College of Science, Kuwait University, Sabah Al Salem University City, P.O. Box 2544, Safat 1320, Kuwait}

\begin{abstract}
\par\noindent
This letter explores how a reinterpretation of the generalized uncertainty principle as an effective variation of Planck’s constant provides a physical explanation for a number of fundamental quantities and couplings. In this context, a running fine structure constant is naturally emergent and the cosmological constant problem is solved, yielding a novel connection between gravitation and quantum field theories. The model could potentially clarify the recent experimental observations by the DESI Collaboration that could imply a fading of dark energy over time. When applied to quantum systems and their characteristic length scales, a simple geometric relationship between energy and entropy is disclosed. Lastly, a mass-radius relation for both quantum and classical systems reveals a phase transition-like behaviour similar to thermodynamical systems, which we speculate to be a consequence of topological defects in the universe.
\end{abstract}

\maketitle

\section{Introduction}
\par\noindent
Various approaches to quantum gravity such as string theory, loop quantum gravity, and quantum geometry suggest a generalized form of the uncertainty principle (GUP) that implies the existence of a minimal measurable length. Several forms of the GUP that include non-relativistic and relativistic forms have been proposed \cite{Amati:1988tn,Garay:1994en,Scardigli:1999jh,Brau:1999uv,Kempf:1994su,Maggiore:1993rv,Capozziello:1999wx,Ali:2009zq}, which can be collectively written in the form $:$

\begin{equation}
    \Delta x~ \Delta p ~\geq ~\frac{\hbar}{2}~ f(p,x)~. \label{GUP}
\end{equation}
Phenomenological and experimental implications of the GUP have been investigated in low and high-energy regimes. These include atomic systems \cite{Ali:2011fa, Das:2008kaa}, quantum optical systems \cite{Pikovski:2011zk}, gravitational bar detectors \cite{Marin:2013pga}, gravitational decoherence \cite{Petruzziello:2020wkd, Al-Nasrallah:2021zie}, composite particles \cite{Kumar:2019bnd}, astrophysical systems \cite{Moradpour:2019wpj}, condensed matter systems \cite{Iorio:2017vtw}, and macroscopic harmonic oscillators \cite{Bawaj:2014cda}. Reviews of the GUP, its phenomenology, and its experimental implications can be found in Refs.~\cite{Addazi:2021xuf,Hossenfelder:2012jw}. 

Recently, we proposed a reinterpretation of the GUP using an effective Planck constant \cite{Ali:2022ckm} by absorbing the additional momentum uncertainty dependence, $\hbar^\prime=\hbar f(p,x)$. This implies a generic GUP of the form:
\begin{equation}
    \Delta x~ \Delta p ~\geq ~\frac{\hbar^{\prime}}{2}~. \label{varyh1}
\end{equation}
Previously, in Ref.~\cite{Chang:2001bm}, the GUP was conceptualized as an effective variation of the Planck constant by isolating an invariant phase space with minimal length in the context of Liouville's theorem. Planck constant was proposed to vary with momentum in Ref.~\cite{Adler:1999js}, where the authors introduced a generalized picture of the de Broglie relation and derived a form of generalized uncertainty principle which is similar to the one obtained in string theory. In Ref.~\cite{Ali:2022ckm}, we argued that the charge radii of hadrons/nuclei along with their corresponding masses support the existence of an effective variation of $\hbar$ that suggests a universality of a minimal length in the associated scattering process. We suggested a relation that simulates the fundamental connection between nature constants ($\ell_{P} ~ M_{P}~ c= \hbar,$) by replacing the Planck length $\ell_P$ by the charge radius ($r$) and the Planck mass $M_{P}$ by the mass of the hadron/nuclei ($m$). This relation reads $:$
\begin{eqnarray}
r~m~c = \hbar^{\prime}~, \label{varyh}
\end{eqnarray}
where $r$ is the charge radius of the particle, $m$ is the particle's  relativistic mass {  \footnote{ The particle's relativistic mass $m$  is given by $m= m_0/\sqrt{1-v^2/c^2}$, where $m_0$ is the rest mass and $v$ is the particle's speed.}, $c$ is the speed of light and $\hbar^{\prime}$ is the effective Planck constant. Here, $c$ is a constant to maintain consistency with the theory of relativity. It was further shown in \cite{Ali:2022ckm} that applying Eq. (\ref{varyh}) to a variety of hadronic particles, as well as to larger nuclei, a clear trend in the effective $\hbar^\prime$ was apparent. It was suggested this effective variation of Planck constant might be related to the timeless state of the universe \cite{Ali:2021ela}. In this letter, we propose a connection between Eq. (\ref{varyh}) and the universal Bekenstein bound \cite{Bekenstein:1980jp} (Sec.~\ref{sec:bekenstein}). Furthermore, we demonstrate that in the case of the electron, a clear connection arises with the value of fine structure constant (Sec.\ref{sec:finestructure}). In this sense, the effective variation of Planck constant is interpreted as a running coupling. We explain the value of the cosmological constant \cite{Ali:2022ulp} (Sec.~\ref{sec:cosmologicalconstant}). In addition, we investigate the conceptual connection between the effective Planck constant and the second law of thermodynamics (Sec.~\ref{sec:secondlaw}). Next, we study the conceptual connection between our formula and both the de Broglie and Compton wavelengths (Sec.~\ref{sec:quantummetric}). Lastly, we present a graphical study of the mass-radius relation, which shows phase transition behavior that may be a consequence of topological defects in the universe. (Sec.~\ref{sec:massradius}). 

\section{The Universal Bekenstein Bound}
\label{sec:bekenstein}
\noindent
The Bekenstein bound is defined as the maximal amount of information contained in a physical system. That is, if a physical system has finite energy and is contained in a finite space, it must be described by a finite amount of information \cite{Bekenstein:1980jp}. Formally, the bound can be written
\begin{eqnarray}
S \leq \frac{2 \pi k_B r E}{\hbar c}~,
\end{eqnarray}
where $S$ is the entropy of the physical system, $k_B$ is the Boltzmann constant, $r$ is the radius of a sphere that encloses the physical system, and $E$ is its energy. Replacing $E=m c^2${  \footnote{ The quantity $m$ is the relativistic mass as defined in footnote 1.}, this becomes
\begin{eqnarray}
S \leq  ~2 \pi~ k_B~  \left(\frac{rmc}{\mathbf{\hbar} } \right)\label{BB}~.
\end{eqnarray}
\par\noindent
One may wonder what happens for the massless particles such as photons. In this case, we get $E= cp$ instead of $E=mc^2$. It is noteworthy that Eq. (\ref{varyh}) is naturally included in the above inequality. Therefore, the bound can be rewritten as  
\begin{eqnarray}
S &\leq& 2 \pi~ k_B \frac{\hbar^\prime}{\hbar}~.
\end{eqnarray}
Replacing the thermodynamic entropy $S$ with the Shannon entropy $H$ \cite{ben-naim},
\begin{eqnarray}
S= k_B H \ln{2}~.
\end{eqnarray}
where $H$ signifies the Shannon entropy, calculated in terms of the number of bits embedded in the quantum states inside the sphere. The $ln 2$ factor comes into play because we interpret information as the base-2 logarithm of the total number of quantum states \cite{Tipler:2005bzt}. We can rewrite the bound as
\begin{eqnarray}
H &\leq& \frac{2 \pi}{\ln{2}}~ \frac{\hbar^\prime}{\hbar}~.
\end{eqnarray}
This is the maximal amount of information required to perfectly describe a physical object up to the quantum level \cite{Bekenstein:1980jp}. 
Effectively, Eq. (\ref{varyh}) introduces a novel way to merge the universal Bekenstein bound with quantum field theory (QFT), through considering the effective Planck constant $\hbar^{\prime}$ for every physical object. The relationship $mrc= \hbar^{\prime}$ corresponds intriguingly to Bekenstein's bound, a universal limit applicable to any physical system for its complete description at the quantum level. Stemming from gravitational insights, Bekenstein's bound serves as a natural cutoff that varies based on the specific physical system being studied. The fact that it behaves like a natural limit led us to propose a connection with renormalization in QFT. Renormalization is a technique employed in QFT to accurately describe physical systems at the quantum level. This is commonly achieved by presuming a cutoff or employing mathematical techniques that provide a cutoff such as counter terms or dimensional regularization that sets limits in order to get finite values instead of infinities. Having uncovered Bekenstein's bound and its implications, we're now seeing a new original meaning for our equation. This overlap provides compelling context to the interpretation of $mrc= \hbar^{\prime}$ as a potential gravitational explanation for renormalization in physics. We expand on this connection in the following sections.


\section{{A running fine structure coupling}}
\label{sec:finestructure}
\noindent
The fine structure constant $\alpha$ describes the fundamental coupling between two electrically charged particles. It is one of the most accurately measured quantities in physics, with a current experimental value of $\alpha=0.0072973525693$ at a precision of $8.1 \times 10^{-11}$  \cite{ParticleDataGroup:2020ssz}.  Originally conceived as a coupling that gauges the basic electromagnetic interaction between the electron and proton, the fine structure constant can be viewed as a quantization of the energy distribution of electrons in the atom. Fundamentally, atomic structure is an artifact of quantum mechanics. As such, we posit that the effective Planck constant for the electron should explain the value of $\alpha$. Using the electron mass $m_e = 9.1093837\times 10^{-31}$ kg  and its classical radius $r_e = 2.8179403262\times 10^{-15}$ \cite{ParticleDataGroup:2020ssz}, we find that effective Planck constant to be
\begin{eqnarray} \label{heff1}
\hbar_{e}^{\prime}= m_{e} r_{e} c= 0.007297352571~ \hbar=\alpha^{\prime}_e \hbar ~,
\end{eqnarray}
where $\alpha^{\prime}_e=0.0072973525\underline{\mathbf{710}}$ is the fine structure constant for the electron obtained from our model. This agrees with the experimentally measured value to 10 decimal places ($\alpha=0.0072973525\underline{\mathbf{693}}$). We infer this to imply an additional physical meaning for the effective variation of Planck constant, {\it i.e.}, as a running fine structure coupling that varies with energy scale. In this sense, an effective variation of Planck constant can be interpreted  as a running fine structure constant $\alpha^\prime$ for every object based on its  mass and its classical/charge radius,
\begin{eqnarray}
\hbar^{\prime}= m r c= \alpha^{\prime} \hbar \label{running}~.
\end{eqnarray}
Interpreted as a running interaction coupling, this may also set a connection between our model and renormalization theories \cite{Weinberg:1980wa}. In fact, it has previously been noted by Adler and Santiago in Ref.~\cite{Adler:1999js} that a running coupling can be conceptualized as a varying Planck constant in a modified vertex term. Our model is consistent with these findings. Additionally, Bekenstein proposed that the fine structure constant should vary based on general principles such as covariance, gauge invariance, causality, and time-reversal invariance of electromagnetism \cite{Bekenstein:1982eu}. A further simple cosmological model was proposed based on the variation of the fine structure in Ref.~\cite{Sandvik:2001rv}, and observations that the fine structure varies with gravity were recently reported in Ref.~\cite{Wilczynska:2020rxx}.


\section{The cosmological constant problem}
\label{sec:cosmologicalconstant} 
\noindent
The cosmological constant $\Lambda$ represents the energy density of the vacuum 
$\rho_{\text{vac}}$ \cite{Weinberg:1988cp} through the relation
\begin{equation}
    \rho_{\text{vac}} c^2=\frac{c^4\Lambda}{8 \pi G}~. \label{CCvac}
\end{equation}
Current astrophysical data \cite{ParticleDataGroup:2020ssz} reports a measured value $\Lambda\approx 1.1\times 10^{-52} ~~\text{m}^{-2}$, which yields an observed vacuum energy density of
\begin{eqnarray}
\rho^{\text{obsv}}_{\text{vac}} c^2= 5.3\times 10^{-10} ~~\text{Joules/m$^3$}~ \label{obsv}.
\end{eqnarray}
In QFT, the vacuum energy density $\rho^{\text{QFT}}_{\text{vac}}$ is calculated as the integration of the vacuum fluctuation energies of all momentum states \cite{Weinberg:1988cp}. For massless particles, this is:
\begin{eqnarray}
\rho^{\text{QFT}}_{\text{vac}} c^2&\approx&
\frac{1}{(2\pi \hbar)^3}\int^{P_{Pl}}_0 d^3p ~~ \frac{\hbar \omega_p}{2} \label{Vaccumdensity} \\&=&\frac{1}{(2\pi \hbar)^3}\int^{P_{Pl}}_0 d^3p\; \frac{c\,p}{2}\nonumber\\
&=& \frac{c^7}{16 \pi^2 \hbar G^2} \label{vacuum1}\nonumber\\&=&2.944\times 10^{111}~~ \text{Joules/m$^3$}~. \label{QFT}
\end{eqnarray}

Comparing the theoretical QFT value of the cosmological constant given in Eq.  (\ref{QFT}) with the observed value Eq. (\ref{obsv}), one finds an outrageous order of magnitude discrepancy
\begin{equation}
\frac{\rho^{\text{obsv}}_{\text{vac}}}{\rho^{\text{QFT}}_{\text{vac}}}= 1.8\times10^{-121} \label{CCp}~.
\end{equation}
This is known as the cosmological constant problem. \\

Let us now define an effective Planck constant for the universe using Eq. (\ref{varyh}), 
\begin{eqnarray}
r_{\text{univ}}~m_{\text{univ}}~c = \hbar^{\prime}_{\text{univ}}~. \label{huniv}
\end{eqnarray}
Using the respective measurements $r_{\text{univ}}=1.37 \times 10^{26}$~m and $m_{\text{univ}}=2\times 10^{52}$ kg \cite{beadnell1940mass,davies2008goldilocks,Gott:2003pf}, the above expression evaluates to
\begin{eqnarray}
\hbar^{\prime}_{\text{univ}}= 8.2\times 10^{86} ~\text{J$\cdot$s}=7.82\times 10^{120} \hbar  \label{heffuniv}~.
\end{eqnarray}
The effective Planck constant of the universe may indicate an increase in quantum probabilities with space expansion \cite{Cotler:2022weg}, and may be related to causal non-linear modifications of quantum mechanics \cite{Raizen:2022xlv}.  It may also be interpreted as a relation between the energy density and the count of spacetime degrees of freedom that is accessible to each particle which has been used to solve a cosmological constant problem \cite{Freidel:2023ytq}.

Returning to the basic definition of vacuum energy density in QFT, we replace the Planck constant in Eq. (\ref{Vaccumdensity}) with the  effective Planck constant $\hbar^{\prime}_{\text{univ}}$ calculated in Eq. (\ref{heffuniv}). In this sense, the vacuum energy density $\rho^{\text{QFT}}_{\text{vac}}$ is replaced by the effective vacuum energy density $\rho^{\text{eff}}_{\text{vac}}$, and we find
\begin{eqnarray}
\rho^{\text{eff}}_{\text{vac}} c^2&\approx&
\frac{1}{(2\pi \hbar^{\prime}_{\text{univ}})^3}\int^{P^{\prime}_{Pl}}_0 d^3p ~~ (\frac{1}{2} \hbar^{\prime}_{\text{univ}} \omega_p) \label{Vaccumdensityeff} \\&=&\frac{1}{(2\pi \hbar^{\prime}_{\text{univ}})^3}\int^{P^{\prime}_{Pl}}_0 d^3p \frac{1}{2} (c~p)\nonumber\\
&=& \frac{c^7}{16 \pi^2 \hbar^{\prime}_{\text{univ}} G^2} =\frac{c^7}{16 \pi^2\times 2.514\times 10^{121} \hbar~G^2} \nonumber\\&=&3.76\times 10^{-10}~~ \text{Joules/m$^3$}~,\label{EQFT}
\end{eqnarray}
where we used the effective Planck momentum $P^{\prime}_{Pl}= \sqrt{\hbar^{\prime}_{\text{univ}}c^3/G}$ as the cutoff of the integration. Comparing the predicted value of the effective energy density   $\rho^{\text{eff}}_{\text{vac}}$ with the observed value $\rho^{\text{obsv}}_{\text{vac}}$, we find \cite{Ali:2022ulp}:
\begin{eqnarray}
\frac{\rho^{\text{obsv}}_{\text{vac}}}{\rho^{\text{eff}}_{\text{vac}}}= 1.41~,
\label{ratio_rho)}
\end{eqnarray}
which is close to unity. This removes the huge discrepancy in the order of magnitude found in Eq. (\ref{CCp}), and thus resolves the cosmological constant problem. The corresponding Hubble constant for our computed vacuum energy density Eq. (\ref{EQFT}) can be calculated as


%
%
%
\begin{eqnarray}
\frac{3 (H_{0}^{\rm eff})^2}{c^2} \Omega_{\Lambda}= \frac{8 \pi G \rho^{\text{\rm eff}}_{\text{vac}} c^2}{c^4}~,
\end{eqnarray}
where $\Omega_{\Lambda}= 0.685$ \cite{ParticleDataGroup:2020ssz}. This yields
\begin{eqnarray}
H_{0}^{\rm eff}=1.85 \times 10^{-18} s^{-1}~,
\end{eqnarray} 
which again agrees to the same order of magnitude with the observed value of Hubble's constant, $H^{\rm obs}_{0}=2.18\times 10^{-18} s^{-1}$ \cite{ParticleDataGroup:2020ssz}. As far as we know, this is the only solution of the cosmological constant problem based on observational data, {\it i.e.}, the mass and radius of the universe, without assuming any additional free parameters or extra dimensions. At this point, it is worth to note that comparing the effective value of the Hubble constant  $H^{\rm eff}_{0}$ with the observed value $H_{0}^{\rm obs}$, we find
\begin{eqnarray}
\frac{H_{0}^{\rm obs}}{H_{0}^{\rm eff}} = 1.18~,
\label{ratio_hubble)}
\end{eqnarray}
which is close to unity as anticipated. It is worth mentioning that previous works \cite{Easson:2010av,Easson:2010xf}  established a connection between the accelerated expansion of the Universe and its large-scale Hawking-like temperature in order to derive a robust estimate for the Hubble constant and other cosmological parameters. Our model proposes that the vacuum energy density is contingent on the radius of the universe and indicates that an increase in the universe's radius would lead to a decrease in vacuum energy density. This may align with recent experimental results suggesting a fading of dark energy over time, as observed by the DESI Collaboration \cite{DESI:2024mwx}.

\section{ The Second law of thermodynamics}
\label{sec:secondlaw}
\noindent
In Ref.~\cite{Ali:2022ckm}, a relation was derived between the effective Planck constant, the area-entropy law \cite{Bekenstein:1973ur}, and the von Neumann entropy $S_N$ \cite{Maldacena:2020ady}, {\it i.e.},
\begin{eqnarray}
\hbar~G~\left(1-\frac{S_{\text{N}}}{S}\right)= \hbar^\prime G = \frac{ c^3 A}{4 S}~, \label{SBH1}
\end{eqnarray}
from which we can define the effective Planck constant as
\begin{eqnarray}
\hbar^{\prime}= \hbar \left( 1-\frac{S_N}{S}\right)~. \label{hentropy}
\end{eqnarray}
Here, $S$ is the entropy of a black hole.
The area-entropy law has been applied to several physical systems in condensed matter physics \cite{Eisert:2008ur} as well as cosmology \cite{Cai:2005ra}. Combining the Hawking temperature for a Schwarzschild black hole \cite{Hawking:1975vcx}
\begin{eqnarray}
T_H = \frac{\hbar c^3 }{8 \pi G k_B~m }\label{Hawking}
\end{eqnarray}
with  Eq. (\ref{running}), we find that
\begin{eqnarray}
\alpha^{\prime}=\frac{r c^4}{8 \pi G k_B T_H}~.
\end{eqnarray}
This gives a new connection between the Hawking temperature and the running fine structure coupling that explains why it is varying with energy. It can thus be understood that every physical object has a corresponding Hawking temperature. We note that the fine structure coupling decreases as the temperature/energy increases, which is consistent  with asymptotic freedom \cite{Gross:1973id}. In comparing our derived formula with the one presented by Peskin and Schroeder in Quantum Field Theory \cite{Peskin:257490}, we find a correlation in certain distance regions. However, differences do appear at extremely short distances, where quantum gravity effects become significant, and at large distances where gravity has a dominating influence over Quantum Field Theory. While these disparities are noteworthy, they could suggest that our derived formula may have the potential to provide additional information or perspective. We can combine  Eq. (\ref{varyh}) with both equations Eq. (\ref{hentropy}) and Eq. (\ref{Hawking}) to obtain the relation
\begin{equation}
\left(S-S_N \right) T_H= \frac{r c^4}{8 \pi G k_B} S~, \label{2ndlawA}
\end{equation}
which can be re-written in the form
\begin{equation}
\Delta S~~T_H= \Delta Q~.
\end{equation}
This is clearly reminiscent of the second law of thermodynamics, provided we associate the change in energy with
\begin{eqnarray}
\Delta Q= \frac{r c^4}{8 \pi G k_B} S~.
\end{eqnarray}
This is a further indication that Eq. (\ref{varyh}) is robust and universally applicable. In addition, it establishes a relation between the change in energy, radius and entropy that reveals the link between information and energy to be purely geometric. It is known that the second law of thermodynamics along with the universal Bekenstein bound can reproduce Einstein's field equations \cite{Jacobson:1995ab,Wald:1993nt}. It thus appears that our approach introduces one more step, in addition to the thermodynamic realization of gravity. That is, our model provides an explicit link between gravitation and the effective variation of the Planck constant. We may also conclude that effective variation of Planck constant could be a reinterpretation of effective field theory at finite temperature \cite{Dolan:1973qd}.

\section{ The de Broglie and Compton wavelengths}
\label{sec:quantummetric} 
\par\noindent
The Compton wavelength of a quantum object is defined as 
\begin{eqnarray}
\lambdabar_C= \frac{\hbar}{m c} \label{compton}~.
\end{eqnarray}
Comparing Eq. (\ref{compton}) with Eq. (\ref{varyh}), or Eq. (\ref{running}), we find
\begin{eqnarray}
\lambdabar_C= \frac{r}{\alpha^{\prime}} \label{cr}~,
\end{eqnarray}
which sets a relation between the Compton wavelength, charge/classical radius, and running fine structure coupling of physical objects. Similarly, the de Broglie wavelength for any physical object moving with velocity $v$ is
\begin{eqnarray}
\lambdabar_{dB} = \frac{\hbar}{\gamma~ m~ v}~,
\end{eqnarray}
where $\gamma=1/\sqrt{1-v^2/c^2}$ is the usual Lorentz factor. When we combine de Broglie-relation with  Eq. (\ref{varyh}), we find
\begin{eqnarray}
\frac{v}{c} \gamma~=\frac{r}{ \alpha^{\prime}~ \lambdabar_{dB}} \label{vc}~,
\end{eqnarray}
which ascribes a geometric interpretation to the ratio $\frac{v}{c}$.  A geometric origin of the Compton wavelength was derived in Ref. \cite{Leblanc:2021vhp} by solving the semiclassical equations of motion for a propagating wavepacket. It was found that the Compton wavelength is equivalent to the square root of the quantum spacetime metric $g_{kk}$, {\it i.e.},
\begin{eqnarray}
\sqrt{g_{kk}} = \lambdabar_C \label{qmetric}~.
\end{eqnarray}
Combining Eq. (\ref{qmetric}) with  Eq. (\ref{cr}), we find:
\begin{eqnarray}
\sqrt{g_{kk}}= \frac{\hbar}{\hbar^{\prime}} ~r= \frac{r}{\alpha^{\prime}}~, \label{original}
\end{eqnarray}
or, alternatively, as
\begin{eqnarray}
m~\sqrt{g_{kk}}= m \frac{r}{\alpha^{\prime}}  \label{original}=  \frac{\hbar}{c}~.
\end{eqnarray}
and so the running fine structure coupling is completely determined by the quantum metric and charge radius. This potentially sheds light on the nature of spacetime foam \cite{Wheeler:1955zz,Hawking:1978pog}.



\section{The Mass-Radius Relation and Topological Defects}
\label{sec:massradius}
\par\noindent
\begin{figure}[h]
\includegraphics[scale=.5]{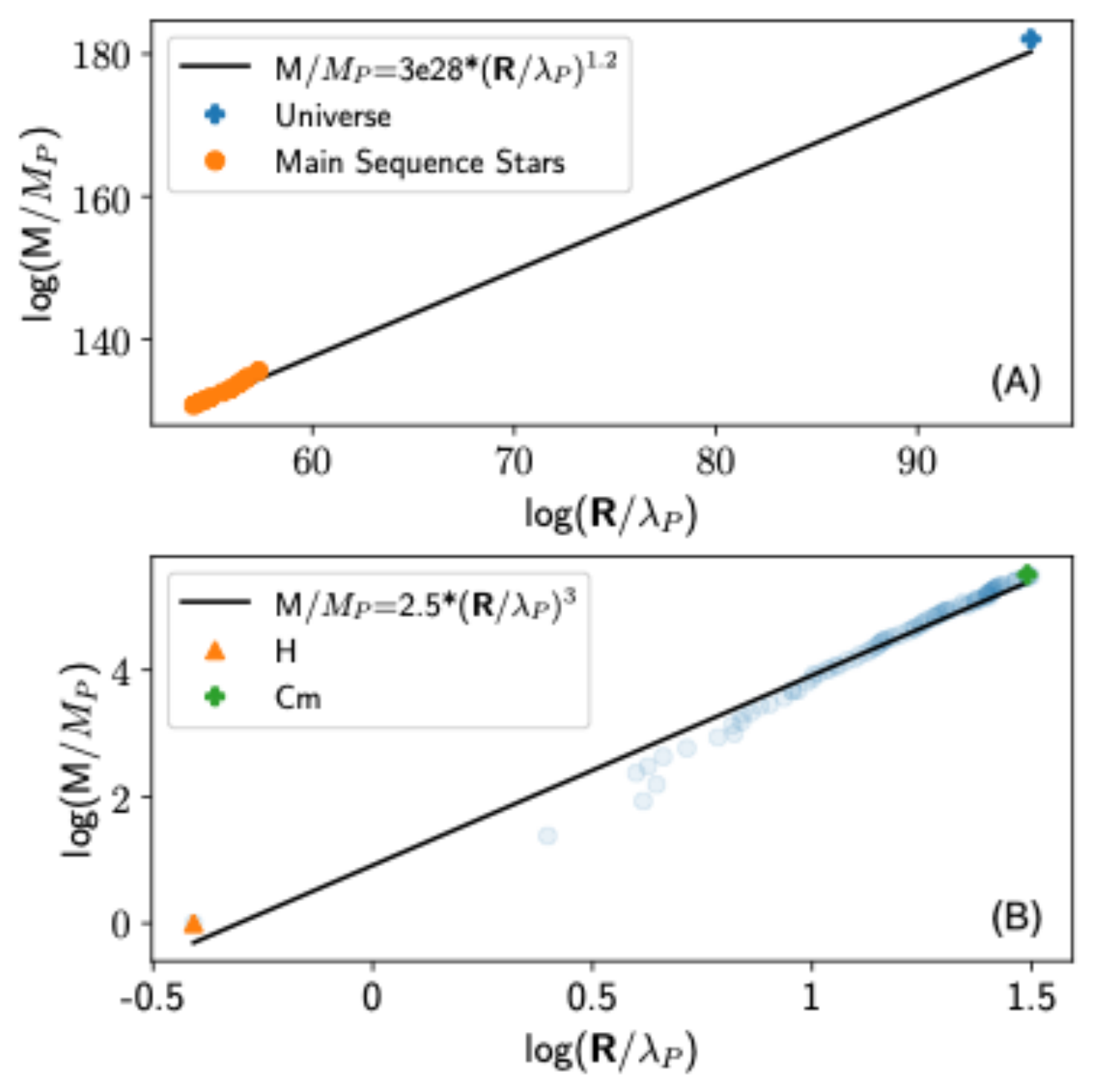}
\caption{Logarithmic plot of mass and radius for (A) main sequence stars and the Universe \cite{Eker123:2018}, and (B) periodic table element \cite{2013ADNDT:2013} normalized by the proton mass and the Compton wavelength of the proton.}
\label{massradius_large}
\end{figure}
\begin{figure}[h]
\includegraphics[scale= 0.4]{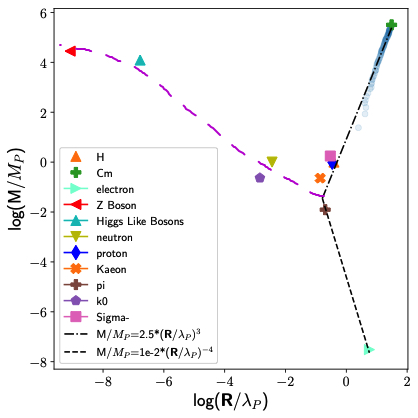}
\caption{Logarithmic plot of mass and radius for fundamental particles \cite{ParticleDataGroup:2020ssz,FRITZSCH2012,BoLehnert12014} and periodic table elements \cite{2013ADNDT:2013}, normalized by the proton mass and the Compton wavelength of the proton.}
\label{massradius_together}
\end{figure}
This implies that the quantum metric is fully determined by the mass of the physical object, which is consistent with general relativity. One can conclude based on Eqs.  (\ref{vc},  and \ref{original}) that the ratio between $v$ and $c$ can be explained by geometric origin. This agrees with timeless state of gravity \cite{Ali:2021ela}. 
Now we explore aspects of the mass-radius relation discussed in the previous sections for a wide range of objects.  Figure~\ref{massradius_large} (A) and (B) \cite{2013ADNDT:2013,Eker123:2018,beadnell1940mass,davies2008goldilocks,Gott:2003pf} shows this for main sequence stars and the Universe (A) and for periodic table elements (B). For these systems, there is a near linear relation between mass and radius. When we look closer at the relation between $M$ and $R$ for fundamental particles, in Figure~\ref{massradius_together}, we find a fluctuation occuring around the linear relation. It is found that some objects have $\hbar^{\prime}/\hbar< 1$, as for the electron, $\pi^{\pm}$ and Z-Bosons \cite{ParticleDataGroup:2020ssz,FRITZSCH2012,BoLehnert12014}, while other objects have $\hbar^{\prime}/\hbar> 1$, like protons and kaons \cite{ParticleDataGroup:2020ssz}. The behavior of $M$-$R$ thus gives information on how the running coupling changes from one physical object to another. The phase transition in the relation,  seen around the pion and kaon cases, could be a result of the existence of topological defects at short distance \cite{kibble1976topology,Bhattacharjee:1991zm} and should be further investigated. Several studies have tackled the relation between the topological defects and the variation of the fine structure \cite{nasseri2005fine}. We notice that the phase transition we obtained in Figure~\ref{massradius_together} is quite similar to the phase transition of the quark-gluon plasma (QGP) \cite{Bhalerao:2014owa}. We speculate the topological defects that describe the QGP \cite{digal1996formation,bhattacharjee1992grand} can be similarly used to describe that of the $M$-$R$ relation. Related to these results, we note that the implications of the mass-radius relation have previously been discussed in the context of dark matter \cite{ianni2022new}.

\section{Conclusion}
\par\noindent
In this letter, we have expanded our analysis of the effective Planck constant derived in Ref.~\cite{Ali:2022ckm} to include more fundamental connections and observational tests. We have shown a conceptual connection between our relation and the universal Bekenstein bound.  When we computed the effective Planck constant for the universe $\hbar^{\prime}_{\text{univ}}$ and considered it in the QFT definition of the vacuum energy density, we obtained a value of effective vacuum energy density that agrees to the same order of magnitude with the observed value. This resolves the cosmological constant problem. We reproduced the fine structure constant when considering the electron case. The effective variation of Planck constant can also be reinterpreted as a running fine structure coupling for every physical object, which may replace renormalization techniques in QFT. The running fine structure coupling is also found to obey asymptotic freedom.  We connect our equation with the Hawking temperature and entropy-area law to reproduce the second law of thermodynamics. Furthermore, we established a connection between charge radius and de Broglie wavelength to introduce a geometric interpretation of the special relativity ratio $\frac{v}{c}$. This facilitates a description of the quantum metric, which is equivalent to the charge radius divided by the running coupling. 

\section*{Compliance with Ethical Standards}
Authors have no conflict of interest to declare. The authors have no competing interests to declare that are relevant to the content of this article.

\section*{Data Availability Statement}

No Data associated in the manuscript.

\section*{Acknowledgments}
J.M. is a KITP Scholar at the Kavli Institute for Theoretical Physics. The KITP Scholars Program is supported in part by the National Science
Foundation under Grant No.~NSF PHY-1748958. The authors are grateful for Caroline Gorham in helping with making the figures in this paper.
\bibliographystyle{utcaps}
\bibliography{ref.bib}{}

\end{document}